\documentclass[apj]{emulateapj}
\usepackage{graphicx}
\usepackage{amssymb}
\usepackage{epstopdf}
\usepackage{color}

\shorttitle{Stripped Tails}
\shortauthors{Kraft et al.~2016}

\begin{document}

\newcommand{\degree}{^o}
\newcommand{\CF}{_{\mathrm{CF}}}
\newcommand{\Hot}{_{\mathrm{hot}}}
\newcommand{\Cold}{_{\mathrm{cold}}}
\newcommand{\Max}{_{\mathrm{max}}}
\newcommand{\ICM}{_{\mathrm{ICM}}}
\newcommand{\KeV}{\,\textrm{keV}}
\newcommand{\Kpc}{\,\textrm{kpc}}
\newcommand{\Kms}{\,\textrm{km}\,\textrm{s}^{-1}}
\newcommand{\K}{\,\mathrm{K}}
\newcommand{\ccm}{\,\mathrm{cm}^{-3}}
\newcommand{\gccm}{\,\mathrm{g}\,\mathrm{cm}^{-3}}
\newcommand{\cmss}{\,\mathrm{cm}\,\mathrm{s}^{-2}}
\newcommand{\Myr}{\,\mathrm{Myr}}
\newcommand{\Gyr}{\,\mathrm{Gyr}}

\newcommand{\etal}{et al.}

\definecolor{rred}{rgb}{1,0,0}
\definecolor{bblue}{rgb}{0,0,1}

\title{Stripped elliptical galaxies as probes of ICM physics: III.  Deep Chandra Observation of NGC 4552 - Measuring the Viscosity of the Intracluster Medium}
\author{R.~P.~Kraft\altaffilmark{1}, E.~Roediger\altaffilmark{2,1} , M. Machacek\altaffilmark{1}, W.~R.~Forman\altaffilmark{1}, P.~E.~J.~Nulsen\altaffilmark{1}, 
C.~Jones\altaffilmark{1},E.~Churazov\altaffilmark{3}, S.~Randall\altaffilmark{1}, Y.~Su\altaffilmark{1}, A.~Sheardown\altaffilmark{2}}
\affil{
\altaffilmark{1}Harvard/Smithsonian Center for Astrophysics, 60 Garden Street, Cambridge, MA 02138, USA
\altaffilmark{2}E. A. Milne Center for Astrophysics, Department of Physics and Mathematics, University of Hull, Hull, HU6 7RX, UK
\altaffilmark{3}MPI f\"{u}r Astrophysik, Karl-Schwarzschild-Str. 1, Garching 85741, Germany
}
\email{rkraft@cfa.harvard.edu}
%\altaffiltext{5}{Visiting Scientist, SAO}

\begin{abstract}

We present results from a deep (200 ks) Chandra observation of the early-type galaxy NGC 4552 (M89) which is falling
into the Virgo cluster.  Previous shallower X-ray observations of this galaxy showed a remnant gas
core, a tail to the South of the galaxy, and twin `horns' attached to the northern edge of the gas core \citep{machacek05a}.
In our deeper data, we detect a diffuse, low surface brightness extension to the previously known tail, and measure the
temperature structure within the tail.  We combine the deep Chandra data with archival XMM-Newton observations to put a strong upper limit on the diffuse
emission of the tail out to a large distance (10$\times$the radius of the remnant core) from the galaxy center.  In our two
previous papers \citep{roediger15a,roediger15b}, we presented the results of hydrodynamical simulations of ram pressure stripping specifically for M89
falling into the Virgo cluster and investigated the effect of ICM viscosity.  In this paper, we compare our deep data with our specifically tailored simulations and
conclude that the observed morphology of the stripped tail in NGC 4552 is most similar to the inviscid models.  We conclude that, to the extent
the transport processes can be simply modeled as a hydrodynamic viscosity, the ICM viscosity is negligible.
More generally, any micro-scale description of the transport processes in the high-$\beta$ plasma of the cluster ICM must be consistent with the efficient mixing
observed in the stripped tail on macroscopic scales.

 \end{abstract}

\maketitle

%****************
\section{Introduction} \label{sec:intro}
%****************
%
The infall of galaxies and small groups into clusters of galaxies can leave a variety of 
visible imprints in both the intracluster medium (ICM) and the corona of the infalling object.  Due to its motion through the ICM,
an infalling galaxy experiences a head wind that exerts a ram pressure onto the galaxy's hot atmosphere causing it to be stripped away 
resulting in the formation of a tail.  Several early-type galaxies in nearby galaxy clusters show this typical head-tail structure indicative of ram-pressure
stripping including NGC 1404 in the Fornax cluster \citep{machacek05a} and NGC 4472 in the Virgo cluster \citep{irwin96,kraft11}.  
One of the best examples of this process is the early-type galaxy M89/NGC 4552 in the Virgo cluster \citep{machacek05b,machacek05c}. By focusing on this galaxy, we will use the 
observed flow patterns of the stripped gas of an infalling galaxy to directly measure the transport coefficients of the ICM. 

The magnitudes of the ICM transport coefficients (i.e. the viscosity and thermal conductivity)  are key unsolved problems in cluster physics.  
The order of magnitude of the effective transport coefficients impact a wide variety of processes critical to the formation
of structure in the Universe such as diffusion, metal transport, sound wave dissipation, and the thermodynamic evolution of galaxies and clusters.
There is growing evidence, however, that the ICM viscosity may be considerably suppressed relative to the Spitzer value.  Recent measurements of
surface brightness fluctuations in the Coma \citep{churazov13} and Perseus \citep{zhuravleva15} clusters measuring the velocity power spectral density (PSD) are
consistent with a turbulent cascade.  The index of the velocity PSD is consistent with a Kolmogorov cascade.  Similarly, application of edge-detection algorithms
on the emission from several clusters demonstrates the presence of arcs of enhanced surface brightness \citep{sanders16}.  Such features
are consistent with KHIs generated by sloshing in an inviscid fluid \citep{roediger13}.  However, the X-ray surface brightness fluctuation measurements have not, as yet,
probed scales smaller than the Spitzer mean free path, and the interpretation of the arcs may be complicated by magnetic effects \citep{werner16,zuhone11}.

In a terrestrial setting, the nature of steady, incompressible flow around a rigid, blunt object depends critically on the viscosity of the fluid.  In particular,
the morphology of the object's wake changes characteristically with Reynolds number.  At Reynolds
numbers of order unity or less (i.e. at large viscosities) the flow is laminar.  As the viscosity decreases and the Reynolds number increases to
$\sim$100, a regular series of cyclonic disturbances appear behind the blunt object called a von Karman
vortex street.  At even lower viscosities ($Re\sim$10$^{4}$), the wake behind the blunt object becomes fully turbulent \citep{landau59}.  
The situation of gas stripping from cluster galaxies is similar to the flow around a rigid, blunt body, but deviates in several important regards (Paper I).  The ICM
is a compressible fluid, the flow is non-steady, the infall is typically supersonic, and the gravitational tidal field
is changing.  The galaxy's atmosphere is not a rigid object, but is also a compressible
fluid that resides in the gravitational potential of the dark matter of the galaxy.  The galaxy's atmosphere will be stripped during the
infall into the cluster, either by the ram-pressure of the infall or by hydrodynamic instabilities, and the stripped galactic gas
will be deposited in the galaxy's wake, potentially giving the wake a tail-like appearance.
Given these complexities, it is not clear how far the simple analogy of incompressible flow
around a rigid, blunt object can be applied to infalling galaxies.  In particular, the effect that viscosity will have on 
the visible appearance of the tail or wake cannot be easily estimated via comparison with the simple terrestrial scenario.
Estimates of the stripping rate of gas in the turbulent and laminar regimes have been made \citep{nulsen82}, but there has 
been no systematic study of the flow patterns in the wake of an infalling galaxy.

Therefore, in the first paper of this series \citet[Paper I hereafter]{roediger15a} we investigated the nature of stripped gas `tails' with 
hydrodynamical simulations that follow the infall of M89
into the Virgo cluster and thus include a realistic stripping history for M89.  
We showed that the blunt body analogy can be applied with a few important modifications.
We distinguish between the galaxy's \textit{remnant tail} and its \textit{wake} 
(see Fig.~3 in paper I).   We demonstrated that the downstream part of a stripped atmosphere is shielded from the ICM and can be largely retained by the 
galaxy up to or beyond pericenter passage. 
This `remnant tail' is part of the remaining 
atmosphere. The flow of ICM \textit{around} the remaining atmosphere is similar to the flow around a solid body. 
The wake of the remaining atmosphere is filled with both stripped galactic gas and 
ICM, and it is only in this region that these two fluids can mix. 
The near part of this wake is a deadwater region which extends roughly one or two  atmosphere lengths in the downstream direction. 
In the rest frame of the galaxy, the velocity in the 
deadwater region is very small or even directed towards the galaxy. 
In our second paper, \citet[paper II hereafter]{roediger15b}, Paper II hereafter) we investigated the impact of an isotropic ICM viscosity on the galaxy's tail and wake. 
If the ICM viscosity is a significant fraction of the Spitzer value, small scale-length Kelvin-Helmholtz instabilities
(KHIs) and mixing will be suppressed in the wake, which will take
the appearance of a long, cool tail.  On the other hand, if the viscosity of the ICM is significantly suppressed relative to the
Spitzer value (i.e. the Reynolds number, Re, number is large), small scale KHIs will rapidly mix the ICM gas with the stripped galaxy gas,
the temperature of the wake will rapidly approach that of the ambient ICM, and the wake will have low gas density and thus be faint.

In this paper, the third of the series, we compare new deep Chandra observations and an archival XMM observation of 
NGC 4552 (M89) to simulations to determine its infall history into 
Virgo and to directly constrain the viscosity of the Virgo ICM. 
NGC 4552  is a large early-type galaxy 1$^{\circ}$ (350 kpc) to the east of M87 in the Virgo cluster (D=17 Mpc).  
\citet{machacek05b, machacek05c} reported
the presence of a remnant core using a 50 ks Chandra observation, and a gas tail bending to the southeast of
the galaxy.  Based on the pressure jump between the remnant core and the ambient Virgo cluster ICM, they estimated an infall
velocity of 1680 km s$^{-1}$ assuming that M89 lies at the same distance as M87, near the pericenter of its orbit.
We measure the temperature and density of the tail and wake as a function of distance from the nucleus. Comparing our measurements with specifically tailored 
viscous and inviscid hydrodynamic simulations, we show that the tail and wake properties agree best with inviscid stripping, thus indicating strong suppression of 
isotropic viscosity of the ICM.  We compare our data with two sets of simulations, one created with a viscosity of 10\% of the Spitzer value and a second with
only numerical viscosity (effectively $\sim$10$^{-4}\times$Spitzer).  Given the limited quality of the data we are only able to grossly distinguish the cases
when the viscosity is a significant fraction of the Spitzer value or when it is suppressed by several orders of magnitude.  No finer gradations are presently
possible.

This paper is organized as follows.  We present a summary of the data and data preparation in Section 2.  Section 3 contains a
discussion of the analysis of the data.  In section 4 we present our simulations and compare our numerical results with
the observations.  We end with a summary and brief conclusion in Section 5.  We assume a distance of 17 Mpc to
NGC 4552, consistent with the surface brightness fluctuation measurement of the distance to subcluster A of the Virgo cluster
which contains NGC 4552 \citep{tonry01}.  All spectral fits are absorbed by the Galactic column of 2.56$\times$10$^{20}$ cm$^{-2}$ \citep{dickey90}
and all quoted uncertainties are at 90\% confidence for one free parameter unless otherwise stated.

%************
\section{Data} \label{sec:data}
%************

NGC 4552 has been observed four times with Chandra (OBSIDS 2072, 13985, 14358, 14359), once in AO-3 and
three times in AO13 giving a total observation time of $\sim$203 ks.  
All four data sets were reprocessed using CIAO 4.7 with the most up to date gain and efficiency calibrations as of that release.  The sub-pixel
randomization was removed, and all visible point sources were removed.   The event data were projected to
a common reference point at the nucleus and merged to create images, and all imaging analysis done on the
combined data set.  Spectral extraction of both source and background regions was done
for each OBSID individually, however.
We examined the four datasets for periods of high background and found little background flaring.
The total good time after removal of background flares is 201.4 ks.

The dominant background relevant for our spectral study of the galaxy's tail is both emission from the ambient 
Virgo cluster gas and the internal particle background.  We therefore use local
backgrounds in all spectral fits.  Dark sky background alone underestimates the total background because of the
proximity (350 kpc) of NGC 4552 to the center of the Virgo cluster.  NGC 4552 is sufficiently far ($\sim$1$^\circ$) from M87 so that there is no significant 
gradient in the cluster emission across the field of one ACIS chip.   The surface density of counts in the tails is sufficiently above background in all cases of
interest that our results and conclusions are insensitive to the specific region chosen for background subtraction.
 
%FFFFFFFFF
\begin{figure*}
\begin{center}
\includegraphics[trim= 0 0 0 0,clip,angle=0,width=0.49\textwidth]{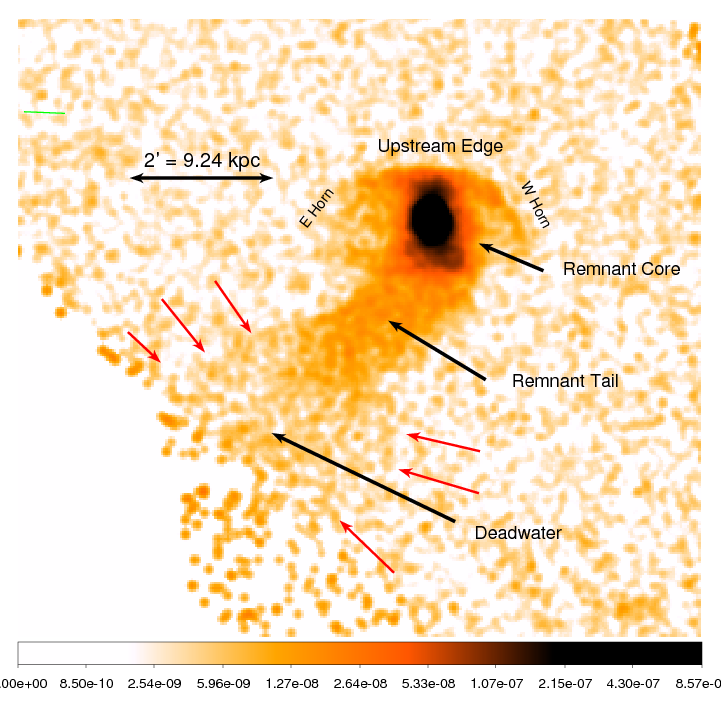}
\includegraphics[trim= 0 0 0 0,clip,angle=0,width=0.49\textwidth]{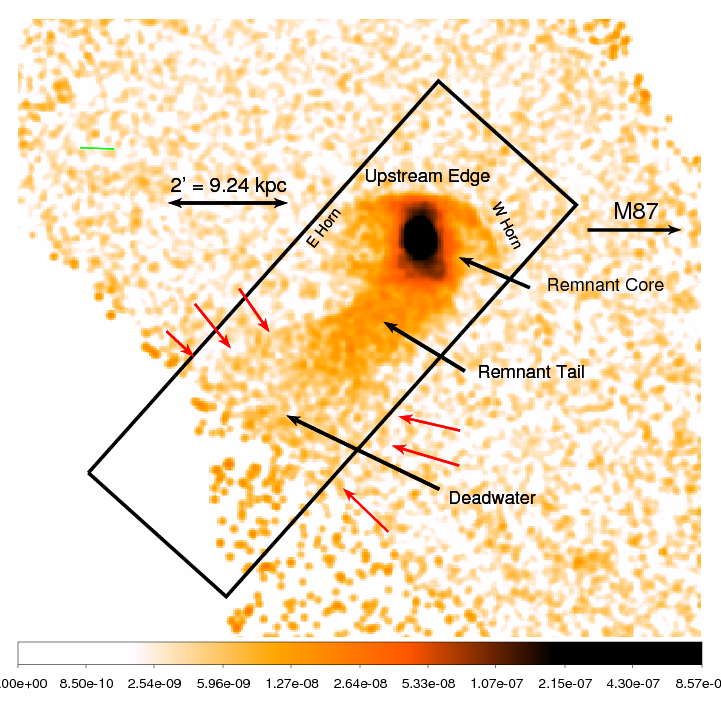}

\end{center}
\caption{Exposure corrected, co-added (200 ks) Chandra/ACIS-S image of M89 in the 0.7-1.1 keV band.  Key features that are labelled include the dense unstripped gas at the
center of the potential (the remnant core), the region of unstripped gas from the halo shielded by the remnant core (the remnant tail), and the region of 
stripped gas that is mixing with the ICM (the deadwater region).  The wake lies $\sim$10$'$ to the southeast off the active region of the ACIS detectors.  The red arrows
denote the boundary of the surface brightness enhancement of the deadwater region and shows the flaring between the remnant tail and deadwater (see text for details).
The units of the color bars are phot cm$^{-2}$ s$^{-1}$ arcsec$^{-2}$ in the 0.7-1.1 keV band.}
\label{fig:n4552}
\end{figure*}
%FFFFFFF
 
 We also use an archival XMM-Newton observation of NGC 4552 to put limits on the flux from the stripped tail beyond the ACIS-S FOV.
 NGC 4552 was observed by XMM-Newton on Aug 10, 2003 (OBSID 0141570101) with the Thin filter in Prime Full Window Extended mode.  For simplicity,
 we use only the PN data in our analysis.  All data were reprocessed from the ODF files using XMMSAS version 20131209\_1901-13.5.0 using the most up to date
 gain and efficiency calibrations as of that release.  The livetime of the observation was 32.1 ks.  We filtered the data for periods of high
 background by binning the data in 100 sec bins in the 5-12 keV band and removing any intervals in which the background rate was 
 more than 3$\sigma$ above the mean.
 We also filtered the data by grade and bad pixel criteria.  This left $\sim$16.7 ks of good time in the PN data.  As with the Chandra data,
 we use local background for background subtraction to precisely remove both the internal particle background and emission from the Virgo cluster.

%*****************************
\section{Analysis} \label{sec:analysis}

\begin{figure}
\begin{center}
\includegraphics[width=0.45\textwidth]{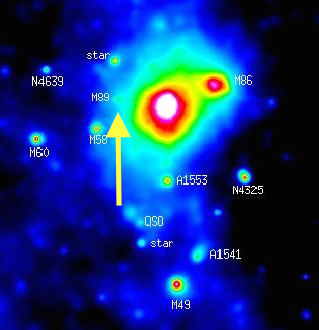}
\end{center}
\caption{ROSAT X-ray image of the Virgo cluster \citep{bohringer95}.  The position of M89 (NGC 4552) is denoted by the yellow arrow.  The projected distance between M87 and M89 is $\sim$350 kpc}
\label{fig:virgo}
\end{figure}

A smoothed, exposure corrected Chandra X-ray image of NGC 4552 in the 0.7-1.1 keV band is shown in Figure~\ref{fig:n4552}.  We choose this narrow band
for the images because the emission is dominated by the Fe L complex of the 0.5 keV galactic gas, 
and thus enhances the visibility of the galaxy gas relative to the hot Virgo cluster gas emission.
The color stretch highlights the tail and wake, for images showing the structure inside the remnant atmosphere
see \citet{machacek05c}.    Detectable X-ray emission in the tail extends to a distance of several times the
radius of the remnant core to at least the edge of the FOV of the Chandra data.
The direction of its tail is curved south to east with a roughly constant width until $\sim$13 kpc from the nucleus where it
flares dramatically and the width increases by a factor of 2 (denoted by the arrows in Figure~\ref{fig:n4552}) while the brightness decreases by a factor of 2.  
The brightest part of the tail ($<$13 kpc from the nucleus) was first reported 
by \citet{machacek05b}, but
the low surface brightness extension is only visible in our deeper data. 
The position of NGC 4552 in the Virgo potential, due east of M87,
is shown in Figure~\ref{fig:virgo}. Direction of its tail indicates north or northwest motion in projection. 

We identify three distinct regions of the galaxy and tail of NGC 4552 visible in the Chandra image
using the terminology described in Papers I and II:  
the remnant core (the brightest region centered
on the optical galaxy consisting of unstripped gas), the remnant tail (the X-ray bright part of the tail just behind the remnant core composed of unstripped gas
that has been shielded from stripping by the remnant core), 
the deadwater region (the fainter region beyond the remnant tail consisting of gas that has been stripped which is 
starting to mix with the Virgo cluster ICM but has no significant velocity
difference from NGC 4552).
The fourth region identified in Papers I and II, the far wake 
(composed of gas well-mixed with the Virgo cluster ICM moving away from NGC 4552 at its infall
velocity), lies well off the ACIS-S3 chip to the southeast.
Extrapolating from the direction of the remnant tail seen in the ACIS observation, we mark its approximate expected position in the XMM image in Figure~\ref{fig:xmm}.    
The far tail is undetected, so this box is only representative of it's possible location to the southeast of the galaxy.  The size and shape of this region
are primarily useful to get the counting statistics correct to estimate the upper limit to its surface brightness.

\begin{figure*}
\begin{center}
\includegraphics[width=0.65\textwidth]{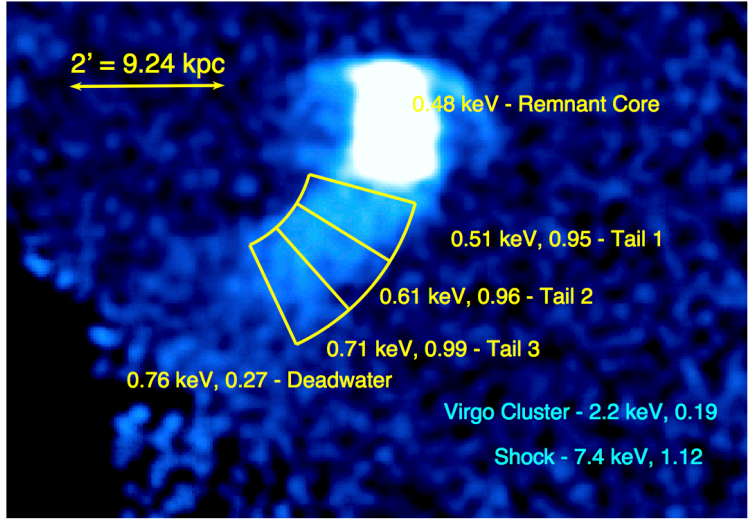}
\end{center}
\caption{Exposure corrected, co-added (200 ks) Chandra/ACIS-S image of NGC 4552 in the 0.7-1.1 keV band.  Three of the four flow regions identified in
Papers I and II are labeled.  The far tail region lies outside the FOV of the ACIS-S3 chip to the southeast.  The two numbers associated with
each region are the temperature (in keV) and pressure (in units of 10$^{-12}$ dyn cm$^{-2}$), respectively.}
\label{fig:regions}
\end{figure*}

\begin{figure*}
\begin{center}
\includegraphics[width=0.65\textwidth]{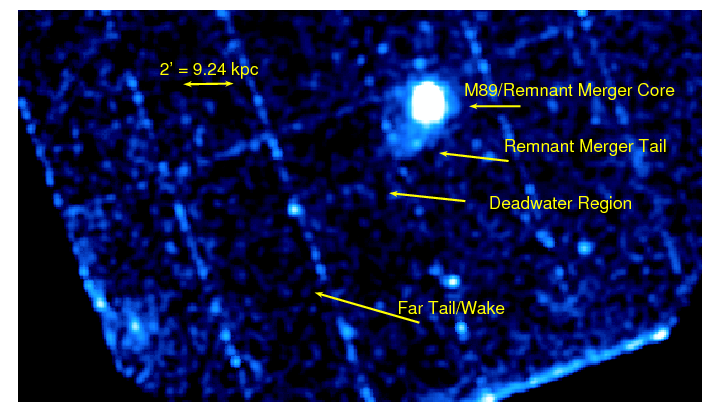}
\end{center}
\caption{Exposure corrected, smooth XMM-Newton/PN camera image of NGC 4552 in the 0.5-1.0 keV band.  The four flow regions identified in
Papers I and II are labeled.  There is no detection of either the deadwater region or the far tail region in the short (18 ks good time) PN observation.
The approximate position of the ACIS-S3 field of view has been overlaid for reference.}
\label{fig:xmm}
\end{figure*}

We fit the spectra of the remnant core, three regions of the remnant tail, and one region of the deadwater region (see Figure~\ref{fig:regions}).
The temperature of the galactic gas still residing in the galaxy potential (i.e. the remnant core) was determined from a small region to the south of the 
nucleus of the galaxy avoiding the region of the AGN 
shock \citep{machacek05c}. A single temperature VAPEC model with Galactic absorption was used in all cases.  The elemental abundances of O, Fe, and Si were 
left as free parameters in the fit of
the remnant core, then fixed at the best fit values for the fits of the tail spectra.  X-ray spectral fitting of diffuse emission from early-type galaxies show some variation
in the abundances of O, Fe, and Si.  
We find abundances of 0.47, 0.29, and 0.30 times the Solar value for O, Si, and Fe, respectively.  The
emission lines of these three elements dominate the spectra at the temperature of interest ($<$1.5 keV), and scaling them all by a constant factor (i.e. using the APEC
model rather than the VAPEC model) would introduce a systematic error in both the density and temperature measurements of the spectral fits.  The abundances
of the other metals is frozen at 0.3 times the Solar value.

%FFFFFFFFF
%FFFFFFF

%FFFFFFFFF
%FFFFFFF

The best fit values of the temperature and uncertainties (90\% confidence) for the remnant tail regions and the remnant core region are shown in 
Table~\ref{tab:tailtemps}.  
The temperature of
the tail of NGC 4552 rises slowly from $\sim$0.48 keV in the region just behind the core to $\sim$0.76 keV in the region where the tail flares.  
We computed the gas density of each
region and the pressure assuming that the infall direction lies in the plane of the sky, that the path length through the tail is the same as the apparent
width of the tail, and that the cool gas in each region has unity filling factor.  
If the infall direction and the tail would be inclined away from the plane of the sky, projection would increase the volume and
therefore decrease the estimate densities and pressures, while small filling factors would decrease the volume (of the cool gas) and thus increase the density.  
In paper I we estimate that the angle of infall relative to the line of sight ranges from 25-45 degrees.  We assume a nominal value of 30 degrees below, but
the uncertainty in the projection angle results in a $\sim$10\% uncertainty in the derived density.
For comparison,
we include the temperatures, densities, and pressures of the ambient and shocked Virgo cluster ICM in Table~\ref{tab:tailtemps}.  
The values for the shocked ICM were determined using the 
Rankine-Hugoniot shock conditions ($\gamma$=5/3) for the Virgo ICM assuming an infall velocity of 1680 km s$^{-1}$, or Mach number=2.2 \citep{machacek05b}.  
This therefore is the the gas temperature, density, and pressure just behind the bow shock.
As we describe below, projection effects will affect our estimates of volumes, densities, etc. to the level of a few tens of percent, but will not change
any of our conclusions.  We discuss the interpretation of these temperature and pressure measurements below in Section 4.

Finally, we examined the archival XMM-Newton observation of NGC 4552 to determine whether we could detect the 
wake beyond the active area of the S3 chip in our deep Chandra data.
An XMM-Newton/PN camera image of NGC 4552 in the 0.5-1.0 keV band is shown in Figure~\ref{fig:xmm}.  
XMM-Newton has significantly more effective area than Chandra and a much larger field of view that is sufficient to cover both the deadwater and 
the far wake regions of the flow.
The background in XMM-Newton is significantly higher than in Chandra making detection of low surface brightness diffuse features difficult, 
and typically a significant
fraction of the observing time is lost to background flares.  
We present only the PN analysis in this paper for simplicity, but our results are not changed 
significantly if we include the MOS data.
There is no detection of either the deadwater region or the far wake in the archival XMM data.  The upper limit to the emission in the 
deadwater region is consistent with our
detection of the deadwater region in the ACIS-S data.  We use the 3$\sigma$ upper limit of the emission from the PN data in the deadwater
and far wake regions to constrain our models of stripping below.

%FFFFFFFFF
%FFFFFFF

In addition to the tail, \citet{machacek05b} reported the presence of `horns' attached to the front of the remnant core (shown in Figure~\ref{fig:n4552}).
Our deep Chandra observation permits us to better map the morphology of the horns and the region between the horns.  We detect faint diffuse X-ray emission
between both horns and the remnant core, which we term the E and W pockets.  
We fit absorbed APEC models to the emission from both horns and from the diffuse emission between the horns.  The best fit temperatures
of the eastern and western horns are 0.50$\pm$0.08 and 0.58$\pm$0.07, respectively.  
Our temperature for the horns is consistent with that found in \citet{machacek05b}.   The temperature of the diffuse emission between the horns and
the remnant core is consistent with these, although with large error bars due to the low surface brightness.
The horns, the pockets, and the remnant core all have similar temperatures, thus indicating a common origin in the remnant core.  
The relative faintness of the emission between the horns and the core
imply that this gas is either mixed with the shock-heated ICM, or fills a thin shell (i.e. a bubble) or a filament around ICM gas.  
We speculate on the nature of these horns and their relation to the stripped tail below.  Detailed analysis and interpretation of these features, 
as well as discussion of the AGN outburst, will be presented in a future paper.

\begin{table}
\begin{center}
\begin{tabular}{|c|l|l|l|}\hline
Region & Temp (keV) & $n_H$  & P  \\ \hline\hline
\multicolumn{4}{|c|}{NGC 4552} \\ \hline
Core & 0.48$\pm$0.03 &  & \\ \hline
Tail1  &  0.51$\pm$0.04 & 5.3 & 9.5 \\ \hline
Tail2  &  0.61$\pm$0.03 & 4.5 & 9.6 \\ \hline
Tail3  &  0.71$\pm$0.05 & 4.0 & 9.9 \\ \hline
Deadwater  &  0.76$\pm$0.10 & 0.98 & 2.7 \\ \hline
Cluster & 2.2 & 0.25 & 1.9 \\ \hline
Shock &  7.4 & 0.43 & 11.2 \\ \hline
\end{tabular}
\end{center}
\caption{Temperatures (keV), hydrogen densities (10$^{-3}$ cm$^{-3}$), and pressures (10$^{-12}$ dyn cm$^{-2}$) for the NGC 4552 core, three regions of the 
remnant tail, one of the deadwater region, and the ambient cluster ICM.  
See text for complete description.}
\label{tab:tailtemps}
\end{table}

%********************

\section{Stripping history, geometry, and the viscosity of the ICM}\label{sec:visc}

\subsection{Identification of remnant tail, wake, and flow patterns for M89}

In our previous papers, we modeled the viscous and inviscid stripping of the infall of NGC 4552 into the Virgo cluster (Papers I and II).  We now compare
the observational data with our models.  In both the viscous and inviscid cases, cool gas from the galaxy is stripped from the infalling galaxy tracing the wake.
We identified four distinct regions of the flow:  the remnant core, the remnant tail, the deadwater (or separation) region, and the far wake.  
These four regions are identified in Figure~3 of Paper I (see also Figures~\ref{fig:n4552} and~\ref{fig:xmm})  
The remnant core, often termed the
remnant merger core in the observational literature, is the unstripped galaxy gas that resides in the dark matter potential.  
The remnant tail is the downstream residual of the more extended gaseous
corona that remains in the gravitational potential of the infalling galaxy shielded from the cluster wind behind the remnant core.  
This remnant tail would typically have similar brightness to the remnant core and 
have similar gas temperature.  
As we described in detail in Papers I and II, the appearance of the remnant
tail will depend more strongly on factors other than the viscosity of the ICM, including the original extent of the gaseous
corona, the path the galaxy takes on the infall into the cluster,  and present position of the stripped galaxy.  That is, the {\it integrated} dynamical history plays
a central role in the present morphological appearance of the gas.
The deadwater region behind the remnant tail is the downstream stagnation region occupied with stripped gas that has little or no velocity relative to the 
remnant core.    Finally, the far wake contains stripped gas mixed with the ICM that
is moving away from the remnant core (in the frame of the galaxy) at roughly the infall velocity.

We identify those expected flow regions for M89 in Figures~\ref{fig:regions} and \ref{fig:xmm}. 
The temperature and brightness of the tail regions RT1-3 strongly suggests that this is indeed the remnant tail. 
The reduced brightness of the region we labelled deadwater region indicates that it is the onset of the wake.  We label it deadwater because the velocity
vectors of the flow in this region should be small relative to the motion of the remnant core.  The ability to distinguish between the
remnant tail and the deadwater regions is important to differentiate between viscous and inviscid simulations.  As we describe in detail below, there is no
possibility that the region we identify as the remnant tail is actually the deadwater region given our underlying assumptions in the modeling.  We can
immediately discount our two compact models since neither have this tail, and if the tail is the deadwater region in the extended atmosphere, viscous case,
it must consist of stripped gas that is not mixing with the ICM.  We should have easily detected the extended deadwater region (which should span
roughly 10$\times$R$_{gal}$ and even the far tail) in the short XMM observation.  To be clear, within our model assumptions, there is no
possibility that the region we identify as the remnant tail is actually the deadwater region.

The pressure of the three tail regions (see Figure~\ref{fig:regions}) we identify as the remnant tail are somewhat lower than the shock-heated ICM behind the bow shock but
roughly a factor of 5 larger than the pressure in the unshocked Virgo region, and are closer to the
estimate of the shocked Virgo cluster gas pressure than that of the unshocked Virgo ICM.  The pressure of the deadwater region
is roughly $\sim$50\% larger than the unshocked Virgo cluster ICM pressure.
Our simulations show that our calculation of the pressure of the
shocked Virgo cluster gas is only appropriate in the region between the bow shock and the contact discontinuity directly in front of the
remnant core (Paper I).  The `cocoon' of shocked gas that surrounds the stripped tail steadily decreases in pressure due to expansion behind the 
bow shock.  At a distance of $\sim$5 core radii, the external pressure of the gas around the tail in our simulations is only a few tens of 
percent higher than that of the unshocked ICM.  Our simulations (Figure 11 of paper II) occasionally show sharp surface brightness discontinuities between
the remnant tail and the onset of the deadwater region in the inviscid case, similar to what we observe.  The apparent large jump in
density and pressure from region RT3 to the deadwater is probably due in part to assuming unity filling factor in the deadwater.
If the gas is starting to mix efficiently here, we underestimate filling factor of the cold gas in the deadwater region, and
therefore underestimated both the density and pressure of this region.  

\subsection{Tail/wake brightness and temperature as indicators of ICM viscosity}

The efficiency of mixing in the wake will depend critically on the viscosity of the ICM.
Generally, if the viscosity of the ICM is a significant fraction of the Spitzer value, all but the largest scale-length Kelvin-Helmholtz instabilities are
suppressed and the galaxy gas will not efficiently mix with the cluster ICM.  In this case, a long, cool, X-ray bright wake will be formed because the galactic gas remains cooler and denser than the ambient ICM.
On the other hand, if the viscosity is low,
the two gases will mix efficiently and the wake will be practically invisible because its density and temperature quickly approach the ambient ICM temperature.
One might therefore naively expect that the presence of the X-ray bright, cool tail (regions RT1-RT3 in Table~\ref{tab:tailtemps})
in NGC 4552 indicates viscous stripping.  As we argue below, the tail in regions RT1-3 is the remnant tail and not the wake. The remnant tail is not
stripped gas, but gas that is bound in the potential of the galaxy shielded by the remnant core.
We identify the low surface brightness region behind the tail with the onset of the deadwater region from simulations.
The deadwater region extends beyond the FOV of the ACIS-S3 chip, and the far wake lies well beyond that.  
Both the deadwater region and the far wake are contained in the XMM-PN FOV, and as described above are not
detected by XMM.

\begin{figure*}
\includegraphics[trim= 0 0 0 0,clip,angle=0,width=0.49\textwidth]{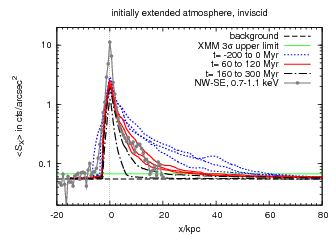}
\includegraphics[trim= 0 0 0 0,clip,angle=0,width=0.49\textwidth]{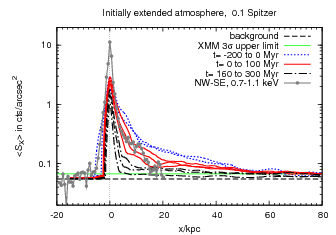}
\includegraphics[trim= 0 0 0 0,clip,angle=0,width=0.49\textwidth]{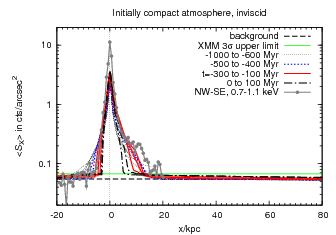}
\includegraphics[trim= 0 0 0 0,clip,angle=0,width=0.49\textwidth]{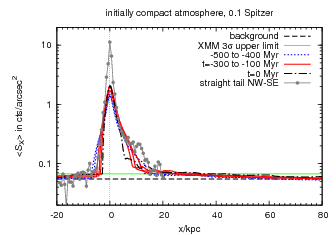}

\caption{Tail brightness profiles for simulation runs and timesteps as labelled in each panel. Time t=0 is pericenter passage.
The profiles for the simulations are averaged across a box 
around the orbit with a width of twice the upstream radius of the atmosphere.  We color-coded profiles from time steps with 
too bright a tail in blue (or too large an upstream radius), with 
tail of approximately the correct brightness in red, and timesteps with too short a remnant tail in black. 
Each color spans a narrow range of time steps to demonstrate the approximate rate at which the surface brightness temporally evolves.
The grey dots are the observed tail brightness extracted from the region shown in 
Fig.~\ref{fig:n4552}.  The XMM 3$\sigma$ upper limit is shown as the green dashed line.  This
limit is estimated from a region that is 12.5 kpc wide and 34.5 kpc long and starts 20 kpc behind the nucleus.  The line extends the
length of the plots for readability.
\label{fig:Xray_tailprofile}}
\end{figure*}

Mixing of the stripped galactic gas and the ICM should occur in the wake unless it is suppressed by, e.g., viscosity; hence the wake is the location to look for 
mixing or its suppression.
We quantitatively compare four specific models to the NGC 4552 data:  viscous (10\% of the Spitzer value) and inviscid ICM, and initially 
extended and initially compact galaxy atmospheres.  
As described in Paper II,
the initial gas density in the galaxy halo prior to infall is unknown, and we bracket our models with two plausible cases with initially extended and initially compact
gas halos.
In Fig.~\ref{fig:Xray_tailprofile} we plot brightness profiles through the galaxy and along its tail and wake for each of the four 
different simulations at eight different time steps around pericenter passage, 
and overplot the observed surface brightness profile for NGC 4552.  Sample images from the simulations that were used to create these profiles are 
shown in Figures~9 and~10 of Paper II.
The 3$\sigma$ upper limit on the emission from the deadwater and far wake regions measured from the XMM-Newton
data is also over plotted in Figure~\ref{fig:Xray_tailprofile}.  The bright peak at the center of the surface brightness profile is due to the AGN.

We distinguish amongst these four models by two observational features.  First, the presence of the remnant tail provides a good estimate of the presence or
absence of an initial extended halo.  Viscosity cannot be distinguished in the remnant tail because that region
contains unstripped galactic gas, which has had no chance yet to mix regardless of the viscosity. 
The two models with initially compact atmospheres, both viscous and inviscid, are ruled out because the remnant tail is 
too short at all times.  The observed
length and brightness of the remnant tail demonstrates that the initially extended atmosphere is the more correct description.  
This model also agrees better with observations of galaxies that still own unstripped atmospheres (see paper I)

Second, the surface brightness of the gas in the deadwater and far wake regions provide a strong constraint on the viscosity.
We rule out the viscous stripping/extended
atmosphere scenario because of the strong limits placed by the non-detection of the far wake in the XMM-PN data 
(shown as the green line in Figure~\ref{fig:Xray_tailprofile}.  That is,
the stripped tail in the viscous, extended corona scenario should have been detected in the XMM-Newton data (at $>$6$\sigma$) if present.
We could also have detected the deadwater region in the viscous, compact atmosphere scenario in the XMM-Newton observation if this were the
correct description:  but we do not.
Therefore, {\it we conclude that the model that best matches our
data is the initially extended atmosphere, inviscid stripping, and that the stripped gas from the galaxy is well mixed with the Virgo cluster ICM in the wake.}
To the extent that we can model the underlying transport processes of the Virgo cluster ICM as a viscous, classical fluid on the macroscopic
scale, the effective viscosity of the Virgo cluster ICM must be suppressed by several orders of magnitude below the Spitzer value.

%FFFFFFFFFFFFFF
%FFFFFFFFFFFFFF

To demonstrate the difference between the four scenarios more conclusively, we plot the measured surface brightness profile along with the `best'
model from each of the four scenarios in Figure~\ref{fig:bestplot}.  We define `best' for each model as the time slice for each of the four cases that most
closely follows the observed surface brightness distribution based on visual inspection.  The four model surface brightness profiles are taken
from four slightly different orbital positions, but all within $\sim$100 Myrs of pericenter passage.
The approximate positions of the remnant tail and deadwater region
are labelled in this plot.  In the remnant tail regions, the compact halo models both under-predict the observed surface brightness, thus excluding these two scenarios.
In the deadwater region, the extended halo, viscous scenario predicts an extended cool tail that if present should easily be observed in the short XMM-Newton
observation.  The only remaining scenario is the extended halo,  inviscid case.

%FFFFFFFFFFFFFF
\begin{figure*}
\includegraphics[angle=0,width=\textwidth]{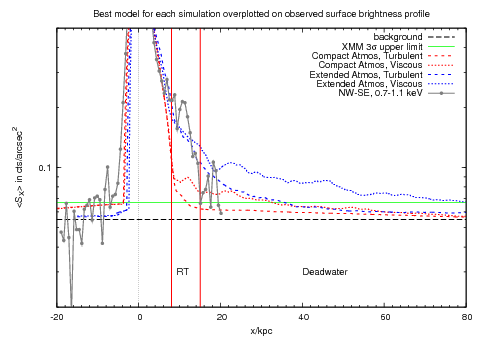}
\caption{Surface brightness profiles for the compact atmosphere viscous and turbulent cases (red dashed lines) and extended atmosphere turbulent
and viscous cases (blue dashed lines).  The observed surface brightness is overplotted (grey points and line) and the background level (black dashed line)
and the 3$\sigma$ XMM-Newton upper limit (solid green line).  The four model curves were taken from Figure~\ref{fig:Xray_tailprofile}, one from
each plot.  The curve from each plot that best matched by visual inspection the observed surface brightness profile was selected.  The compact atmosphere
models lack the remnant tail, and the extended halo, viscous model would produce a long, cool, bright tail of stripped gas at a brightness well above
the limit from the XMM observation.  The only model that fits both of these features is the extended halo, turbulent model. \label{fig:best plot}}
\end{figure*}
%FFFFFFFFFFFFFF

We note that temperatures in the deadwater or far tail regions are not strong diagnostics between viscous or inviscid flows.  Again, in principle, we expect a long, 
cool tail for viscous stripping, and a well mixed flow for inviscid stripping.
Our simulations show that in the viscous case the temperature in the deadwater region and far wake should increase only slowly.  However, even if the
viscosity is low and the mixing is efficient, (i.e. the filling factor of the cold gas is low in the far wake), the emission is almost always dominated by the cold gas 
given the $n^2$ dependence on the volume emissivity.  
Therefore unless the filling fraction of the cold component is negligibly small, the spectra will always be dominated by the cooler component, and a single
temperature fit will be driven toward the lower temperature of the galaxy gas.  
We confirmed with XSPEC simulations that unless the volume filling fraction of the cold gas is less than 5\%, the best-fit temperature of a 
single temperature fit will be close to
the temperature of the cold component.  The quality of the existing data is not sufficiently high for us to be able to reliably estimate the filling factor of the two fluids in the 
tail from a two temperature spectral fit.
The surface brightness of the tail is a much stronger diagnostic of viscosity than the temperature.
Finally, we note that our viscous simulations addressed the case of an isotropic viscosity and our comparison between simulations and 
observation ruled out only this case. The transport processes in the ICM are likely anisotropic due to their alignment with the local magnetic field direction 
\citep{kunz15,zuhone15}, and thus may not be able to suppress KHIs as easily \citep{suzuki13,roediger15b}.  
On the other hand, magnetic fields themselves can potentially suppress KHIs. 
However, {\it the non-detection of the wake strongly limits the presence of any mechanism to reduce mixing in the wake}.

\subsection{Dynamical Constraints and Projection}

We can derive additional constraints on the geometry of infall from comparison of the simulations to the data.
The remnant tail extends $\sim$3 upstream radii towards the downstream direction. The upstream radius is defined as the distance from the galaxy
center to the contact discontinuity in the direction of infall.  Long tails of this brightness exist in our simulations
only for pre- or near-pericenter stripping (Fig.~\ref{fig:Xray_tailprofile} and paper I)
independent of viscosity.   
These remnant tails are, however, eroded after pericenter passage.   
Thus, the presence of the remnant tail in NGC 4552 argues that the galaxy is close to the same distance as the Virgo center,
rather than being much closer to us as indicated by 
optical surface brightness fluctuations (\citealt{mei07}). 
Instead, NGC 4552 is likely close to the pericenter of its orbit and moving close to the plane of the sky.
Additionally, the total infall velocity
is known from the stagnation point analysis \citep{machacek05b}.  Combined with the relative velocity differences 
between NGC 4552 and M87, we conclude that its motion is close to the plane of the sky, on an orbit that is inclined by about $30\degree$,
such that NGC 4552 moves towards us.  
Thus NGC 4552 is likely near pericenter given its velocity.

We confirm that our results are not significantly affected by the inclination by comparing with simulations.  We show simulated X-ray images of the initially extended 
atmosphere modified by inviscid stripping at six different inclinations
to the line of sight in Figure~\ref{fig:Xray_inclination}.  
The observed morphology of the galaxy tail is virtually independent of viewing angle unless the infall is 
nearly along the line of sight (the right-most figure).
These figures do show one interesting effect of viewing angle.  The apparent distance between the bow shock and the stagnation 
point at the head of the galaxy is a strong function
of the viewing angle.  In principle, the distance between the bow shock and the stagnation point is a strong function of the Mach 
number of the shock and could be used as
an independent estimate of velocity.  As seen in these images, this distance is a strong function of inclination angle.  
When the infall direction lies close to the line of sight,
we are observing the shock along its wings where the shock is weaker.  The strongest part of the shock along the primary streamline is hidden by projection.
Additionally, in inviscid stripping too large and inclination projects KHIs from the far side of the contact discontinuity to the apparent stagnation point, as we observe in 
NGC 1404 (Su \etal in press).
Since this is not observed, the inclination of M89's direction of motion must be less than $\sim$45 degrees, consistent with our other estimates and those of \citet{machacek05b} based
on the recessional velocity relative to M87 and the measured infall velocity estimated from the pressure jump across the shock.

\subsection{Direction of infall}

\begin{figure*}
\includegraphics[trim= 0 0 100 0,clip,angle=0,width=0.99\textwidth]{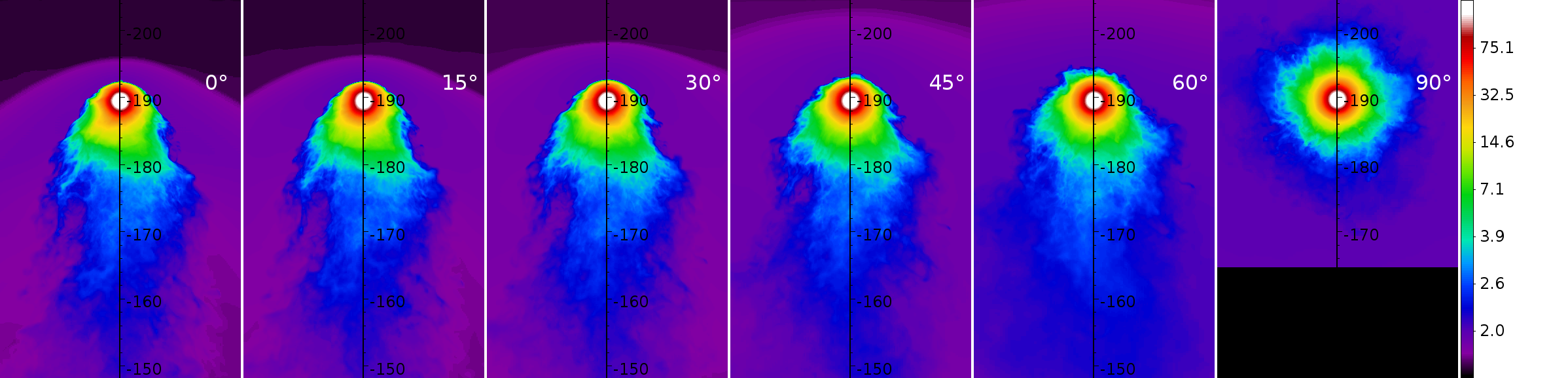}
\caption{X-ray images (0.7-1.1 keV) of a simulated gas-stripped galaxy (surface brightness in arbitrary units but the relative scale is given in the color
bar on the right). Initially extended atmosphere, inviscid stripping, 140 Myr after pericenter 
passage, for different inclinations of the orbital plane (rotation around the horizontal axis, see labels in each panel). The axis is labelled in kpc and shows the distance to pericenter. 
Same as Fig.~16 in paper II, but with  color scale highlighting the projected position of the bow shock.   The visible appearance of the remnant tail and wake is largely independent of the
viewing angle unless it is close to the direction of motion.
\label{fig:Xray_inclination}}
\end{figure*}

NGC 4552's remnant tail extends from the southern boundary of the remnant core, but is distinctly bent to the east as shown in Fig.~\ref{fig:n4552}. 
Such sharp bends in the remnant tail occur occasionally in the simulations (see Fig.~1 of paper II). 
They arise from KHIs that grew to their maximal size while moving along the side of the remnant tail as shown in Fig.~\ref{fig:Xray_inclination}. At this stage, 
they have a sharper upstream interface to the ICM, whereas the brightness on their downstream side fades slowly. This structure is 
matched by the sharp north-eastern 
border of NGC 4552's near tail and its more diffuse south-western border. 
Furthermore, ICM wind fluctuations on scales of a few 10 kpc, i.e., a variable side wind, could bend the tail.   Cluster atmospheres are
dynamic environments, and the assumption that the ICM can modeled with as spherically symmetric and hydrostatic is poor
under close examination \citep{morsony10}.
A further possibility is an ellipticity of the galactic potential, 
which we neglected in our simulations. NGC 4552's near tail bends towards the major axis, i.e., the shallowest potential gradient.

%FFFFFFFFFFFFFF
%FFFFFFFFFFFFFF
%%*******  
\subsection{Horns or KHI rolls}

In our inviscid simulations, KHIs along the sides of the atmosphere and remnant tail are ubiquitous, and somewhat resemble the horns and pockets observed to the side of the NGC 4552 remnant core
(see Figure~\ref{fig:n4552}). 
Such features are rare in our 0.1 Spitzer-viscous simulations once the atmosphere is reduced to its observed size.  In this interpretation, the presence of
the horns and pockets favors a low ICM viscosity. 
Despite these encouraging agreements, our simulations fail to match some subtleties in observed structures. First, the observed horns are twice as long as 
NGC 4552's upstream radius, 
and are attached to its upstream side. In the simulations, such large horns/KHI rolls appear only further downstream along the atmosphere and remnant tail. 
However, inclining the simulated orbit by about 30 degree can shift the horns closer to the upstream edge in projection (Fig.~\ref{fig:Xray_inclination}). 
Then the projected horns in the simulated images  resemble NGC 4552's diffuse eastern horn and the emission on its downstream side, but they do not reproduce the 
bright rim of the western horn. 

As an alternative, we speculate that the horns could be the result of a previous epoch of nuclear activity.
The center of NGC 4552 hosts a young AGN outburst driving a strong shock into the galaxy gas \citep{machacek05c}. 
NGC 4552 could have undergone another earlier AGN outburst which expelled part of its atmosphere and, in connection with the stripping, formed the horns and pockets.
The morphology of the gas in the remnant core is not symmetric due to the ram pressure forces.  An AGN outburst, whether aligned with the infall direction or offset by
some angle, will propagate through the ISM asymmetrically.
Once an AGN shock passes through the contact discontinuity between the galaxy gas and the shock-heated Virgo cluster gas, the remnant core will begin to
slosh in the gravitating potential.  If the direction of the AGN outburst is close to the infall direction, the sloshing will be primarily along this direction as well.
We created some test simulations of this effect and a powerful AGN outburst does induce sloshing, and once the sloshing starts,
large KHI rolls can be stripped off the leading edge of the remnant core.
This mechanism indeed works, but large horns at the front edge are short-lived.  We are investigating this phenomenon systematically with specifically-tailored
hydrodynamical simulations and results will be presented in a future publication.

Finally, we note that we ignored the effects of AGN-induced sloshing in our simulations for modeling the surface brightness profiles to measure the
viscosity.  It is possible that a sufficiently powerful AGN outburst could induces violent sloshing that would temporarily increase the stripping
rate.  In this case the region that we identified as the remnant tail may be the result of this short epoch of violent AGN-induced sloshing.  It is {\it possible} that
there are periods in which the compact atmosphere, viscous case could mimic the features that we see, at least for short periods.  The impact of AGN-induced
sloshing would need to be fully and systematically investigated to assess the importance of this phenomenon.

%**********

\section{Summary and Conclusions}

We report the direct comparison between deep Chandra and archival XMM-Newton observations of the stripped tail of NGC 4552 and 
specifically tailored hydrodynamic simulations of
its infall into the Virgo cluster.  The morphology of the tail in the Chandra image, its temperature and pressure profiles,
and the lack of detection of the deadwater and far wake regions in the XMM-Newton data strongly
favor inviscid stripping.  {\it We conclude that the viscosity of the Virgo cluster ICM is significantly suppressed relative to the isotropic Spitzer value.}
This conclusion would be strengthened with a much deeper XMM-Newton observation centered on the deadwater region and far wake.  Only a small portion of the deadwater
region is located on the ACIS S3 chip in the present observation.  Mapping the brightness and extent of the emission from the far wake with a next 
generation X-ray observatory
such as Athena \citep{nandra13} or X-ray Surveyor \citep{alexey12,gaskin15} would allow for a much more detailed comparison 
between the data and the simulations.  We eagerly the await the results from the scheduled deep Chandra observation of the outskirts of the Coma
cluster (PI:  I. Zhuravleva).  The exposure is sufficiently long and
the gas temperature and density are such that Chandra's resolution is below the Spitzer mean free path.  Hence, one can probe the surface brightness
fluctuation spectrum at the scale of the mean free path to test for a steepening that would be required in any non-magnetized (pure hydro-like)  plasma.

Modeling the underlying transport processes of the ICM as an isotropic viscosity is almost certainly an overly simplistic characterization
of the microscopic physics.  In fact, it is almost certain that MHD and plasma process play a key role in the underlying transport processes \citep{squire16,kunz15},
and that these processes may be determined on scales as small as the proton gyro-radius, which for typical Virgo ICM
parameters would be many orders of magnitude
below the spatial scales we are probing here.  However, whatever these underlying processes are, the ICM apparently behaves as an inviscid fluid 
on macroscopic scales.  The galaxy gas in the wake that has been stripped off the NGC 4552 remnant core is well mixed with the Virgo cluster ICM.
Any viable model of ICM transport processes must incorporate efficient mixing of fluids on macroscopic scales.

A second key result of this work is that context matters.  That is, the appearance of the stripped tail depends critically on its history.  It is not sufficient to observe
the tail of a stripped galaxy and simply use the presence or absence of a cool tail as evidence of viscous or inviscid stripping.  Even for high Re flows (Re $>$ 10000) which
should be fully turbulent, the
stripped galaxy may exhibit a long, cool tail because of a combination of effects including the infall history, the extent of the original gas corona, and the depth of the
central potential.  Each case must be considered as an individual example.  We emphasize that NGC 4552 is a relatively simple case in that the 
size scale of the galaxy is
small compared to the distance to the center of the Virgo cluster.  NGC 4552 has undergone a slow, steady increase in the ram pressure of the 
ICM during its infall, and
the effect of gradients in the pressure of the Virgo ICM are not significant.  

Finally, we make a speculative prediction related to the presence of the horns.  If these features are, in fact, related to a previous epoch of nuclear activity, 
radio bubbles must
have been generated by the outburst.  Once these bubbles crossed the contact discontinuity between the galaxy gas and shocked Virgo cluster ICM, they would have
been rapidly swept up with the flow.  We speculate that there may be relic radio bubbles several tens or hundreds of kpc behind (i.e. to the south or southeast) 
NGC 4552.  
The bubbles may have been shredded by
the strong wind, but given that radio bubbles generally survive in a hydrodynamic sense for tens or hundreds of Myrs in radio galaxies, 
they may have survived the interaction with the Virgo cluster ICM and could have been
advected behind NGC 4552.  Given the likely age of these bubbles (tens or hundreds of Myrs), they may only be detectable in low frequency 
observations due to synchrotron aging and adiabatic expansion.

\section*{Acknowledgements}

This work was supported by NASA grant NAS8-03060 and NAS-GO2-1314X.
E.R.\ acknowledges support by the Priority Programs 1177 ("Witness of Cosmic History") and 1573 ("Physics of the Interstellar Medium") of
the DFG (German Research Foundation), the supercomputing grants NIC 4368 and 5027 at the J\"{u}lich Supercomputing Center, a visiting
scientist fellowship at the Smithsonian Astrophysical Observatory, and the hospitality of the Harvard/Smithsonian Center for Astrophysics in Cambridge, MA.
We would like to thank the anonymous referee for many recommendations that significantly strengthened this paper.

%*******************************************************************
%*************** R E F E R E N C E S *******************************
%*******************************************************************
%
\bibliographystyle{apj}
%\bibliography{library}
%\bibliography{/Users/kraft/mendeley/library}
\end{document}